\documentclass[a4paper, 12pt]{article}

\usepackage[english]{babel}
\usepackage{amsmath}
\usepackage{graphicx}
\usepackage{fullpage}
\usepackage{enumerate}
\usepackage{gensymb}

\title{What  The  Trace  Distance  Security  Criterion  in  Quantum  Key  Distribution Does  And  Does  Not  Guarantee}

\author{Horace P. Yuen\\Department of Electrical Engineering and Computer Science\\Department of Physics and Astronomy\\Northwestern University\\Evanston Il. 60208\\email: yuen@eecs.northwestern.edu}

\date{}

\begin{document}
\linespread{1}
\maketitle
\linespread{2}
\begin{abstract}

Cryptographic security of quantum key distribution is currently based on a trace distance criterion. The widespread misinterpretation of the criterion as failure probability and also its actual scope have been discussed previously. Recently its distinguishability advantage interpretation is re-emphasized as an operational guarantee, and the failure probability misinterpretation is maintained with a further failure
probability per bit interpretation. In this paper we ​explain the basic perpetuating error as a confusion on the correspondence between mathematics and reality.  We note ​​that the assignment of equal a priori probability of 1/2 to the real and ideal situations for distinguishability advantage would not lead to operational guarantee. We explain why operational guarantee in terms of Eve's probabilities of getting various key bits is necessary for security​, and ​why the ​failure probability interpretation misrepresents the security situation.​​ The scope and limits of the trace distance guarantee are summarized.​ It is shown that there would have been no security problem to begin with if the failure probability per bit interpretation has validity.

\end{abstract}

\section{Introduction}
The security of quantum key distribution (QKD) is currently quantified through a trace distance criterion $d$. This criterion $d$ has been misinterpreted [1,2,3] as the ``failure probability" of  the QKD cryptosystem, the probability of its being not ideal to the users. The mechanisms of this error and its security implications have been discussed in some detail [4,5,6] and references cited therein. However, the misconception is still widely maintained with no answer to the points raised a​gainst​ it. The original precise erroneous statement is no longer repeated recently and vague claims​ are made​ instead, which perpetuate the favorable misinterpretation of perfect security with a high probability ​under​ the criterion.

More recently, ref [7] has appeared which further entrenches the misinterpretation as sufficient operational guarantee through a drawn out ``distinguishability advantage" argument adopted from conventional cryptography and first made for QKD in [8]. The ``failure probability" interpretation is maintained without clear demarcation from the previous incorrect one.  A new quantity $d/l$ where $l$ is the total number of bits ever generated by the QKD system but $d$ is the level of one QKD round, is newly introduced [7] as ``failure probability per bit", which would imply quantitative numerical adequacy of the $d$ level even when it is totally inadequate by any account. 

In this paper we would explain the nature of operational guarantee for information theoretic security and show that distinguishability advantage is not operational and does not have the significance attributed to it. It is shown that $d/l$ is not any failure probability per bit, even for just one round as used in [9]. Some comments on distinguishability were previously made [10] but here we focus on the incorrect assumption of equal a priori probability of the real and the ideal situations to the users​ which is a point that has not been brought out before. We also show that there is no operational guarantee from distinguishability even under such assumption, due to the nature of binary hypotheses testing. More fundamentally, we clarify the underlying error of both the failure probability and distinguishability advantage interpretations in their confusion about what reality is referred to by the mathematics.  ​

​Indeed, t​he troubles and misinterpretations surrounding the significance of $d$ are conceptual​ ​and not mathematical issues. They concern the relation of the mathematics to the real world. Such issues often arise in somewhat different ways in engineering as compared to physics. They are sometimes subtle in cryptography, especially in physics-based cryptography. Incorrect interpretation of correct mathematics may lead to totally incorrect conclusions, or very misleading ones, which have happened quite a few times in the QKD literature and will be more fully addressed in a forthcoming book.

​Security proof in cryptography, whatever their practical significance, can only be obtained mathematically and conceptually and not by experiments. The foundation of QKD security and also classical information theoretic security in general must thereby be based on thorough and careful analysis and examination. Cryptographic security is a serious matter and the system designer must make clear \textit{what is and what is not guaranteed} in a real world that is correctly captured by the mathematical representation. This paper tries to present some new simple arguments to bring out the criterion problems that would have been hidden in a treatment focusing on formalism.

In the next section we would point out the irrelevance of distinguishability advantage as an operational security guarantee by itself​, and explain the underlying conceptual error​ which attributes significance to it that it does not have. In section III the meaning of failure probability from $d$ is analyzed and shown to be incorrect when it is made precise, and misleading when left vague in its relation to operational probabilistic security. The requirement of operational guarantee is explained and it is shown how incorrect conclusions are obtained from the failure probability interpretation of $d$ for various important security questions. In section IV the criterion of failure probability per bit is analyzed and shown to be a totally mistaken notion. 

\section{Distinguishability Advantage and A Priori Probabilities} 

The trace distance criterion in QKD is defined by $d\leq \epsilon$, where the trace distance $d$ between the real and the ideal situations to the users is given by the trace norm of the difference between the real situation density operator $\rho_{real}$ and the ideal situation one described by $\rho_{ideal}$,
\begin{equation}
d\equiv \frac{1}{2} ||\rho_{real}-\rho_{ideal}||_1
\end{equation}
We need not be concerned with the exact form of $\rho_{real}$ and $\rho_{ideal}$ [1-3, 5-7], and may just note that in the ideal situation, the QKD generated key $K$ is perfectly random (the uniform random variable $U$) to the attacker Eve. The trace distance criterion is adopted due to the inadequacy of the accessible information criterion [12,13] widely employed previously, that knowing $\log n$ data bits in a known-plaintext attack may reveal the entire $n$-bit $K$. Trace distance is the quantum generalization of the classical statistical distance (variational distance [13], Kolmogorov distance) $\delta(P,Q)$ between two probability distributions $P$ and $Q$,
\begin{equation}
\delta(P,Q)\equiv \frac{1}{2} \sum_i |P_i-Q_i|
\end{equation}
The bound $\delta_E\equiv\delta(P_K, U)\leq \epsilon$ is obtained from (1) and $d\leq \epsilon$, whenever a measurement is made on the system described by $\rho_{real}$ with probability distribution $P_K$ for the $N=2^n$ possible values of the $n$-bit generated key $K$. Hence the guarantee from $d$ is read out entirely through the classical $\delta_E$, and one can equivalently consider the significance of the guarantee $\delta_E\leq \epsilon$. In this paper we would just use $d$ or $\delta_E$ at its own numerical level instead of using its bounded value $\epsilon$, for simpler presentation without any loss of generality.

Recently, the distinguishability advantage interpretation is being presented again formally with incorrect and misleading interpretations and conclusions [7], without answering directly the points raised in [6,14,15] and other previous papers. Perhaps the ``operational meaning" of distinguishability advantage and the ``failure probability per bit" interpretation are the intended answers, which are to be analyzed in this paper. In this section we will deal with the distinguishability issue as follows.

From binary quantum detection theory [16], the probability $P_c$ of correctly deciding between two hypotheses described by density operations $\rho_0$ and $\rho_1$ with a priori probability $p_0$ and $p_1$ can be obtained. Though it seems it has not been given explicitly in the literature except when $p_0=p_1=1/2$, the formula for $P_c$ can be written as
\begin{equation}
P_c=\frac{1}{2} + \frac{1}{2} ||p_0\rho_0-p_1\rho_1||_1
\end{equation}
When $p_0=p_1=1/2$, the trace distance ``distinguishability advantage" term in (3) is just the $d$ of (1) with hypothesis 0 being the real situation and hypothesis 1 the ideal situation. When $p_0\neq p_1$, $d$ is not related to $P_c$ from which the distinguishability advantage interpretation is derived. It is easily seen from (3) that $P_c$ is biased toward hypothesis 0 when $p_0>p_1$, as it intuitively should, and it approaches 1 for hypothesis 0 when $p_0$ approaches 1. It is not known how (1) is related to the distinguishability advantage $P_c-p_0$ when $p_0\neq 1/2$.

Indeed, if it makes sense to assign a priori probability to the real versus the ideal situations in a binary discrimination problem, the a priori probability of the real situation should be 1 and the ideal situation 0. This has been explained in [5,6] that for $d$ or $\delta_E >0$, the probability distribution $P_K$ would not be $U$ with any nonzero probability without adding in further conditions that are far from being provided by $d\leq \epsilon$. To begin with, the discrimination problem has no empirical meaning because we all know we are in the real situation for sure. Furthermore, Eve never cares to make such a discrimination, her aim is to learn about the generated key $K$. The hypothetical game of discriminating between the real and the ideal situations has \textit{no operational security meaning} not only because of these reasons but also because there is no basis to assign any nonzero a priori probability to the ideal situation. On the other hand, the a priori probability affects the quantitative distinguishability advantage a lot as we see above from (3). 

In fact, there are many hypothetical situations, say the ones between the $d=\epsilon$ level and the $d=0$ ideal level. Should we do a multiple hypothesis decision problem on the situation? Why not a binary decision game between two situations with at least one of which less secure than the real one, and conclude from the binary decision that one is in a worse situation than the real one? The inapplicability of the distinguishabiity interpretation of the ideal situation being in effect can be observed from the following consequence (or basis) of (3) for the assumed equal a priori probability,
\begin{equation}
P(ideal|ideal)=\frac{1}{2}+\frac{d}{2},\:\:\:\:\:\:P(ideal|real)=\frac{1}{2}-\frac{d}{2}
\end{equation}
Why would the ideal situation have such a high probability close to 1/2 for \textit{any} $d\ll 1$? It is because $p_1$ is taken to be 1/2 to begin with, but as noted above it is actually 0. It is clear that the operational meaning from $d$ has to be derived \textit{without} assigning a priori probability to the ideal situation, especially not one as big as $1/2$. Such operational meaning of $d$ could be and have been provided in [6] and references cited therein.

One major problem of using such distinguishability advantage argument is that it quickly becomes in one's mind an indistinguishability statement, that the real situation and the ideal situation are only distinguishable with probability $d$, which is in fact identical to the \textit{original m​is​interpretation} of $d$ as the ``maximum failure probability" with ``failure" meaning the cryptosystem is different from an ideal one [1,2,3]. This interpretation of quantitative indistinguishability as failure probability is common and stated explicitly in [17, p.3]. It has been used as an alternative derivation of the wrong interpretation. Among other problems such indistinguishability confuses the meaning of a mathematical statement with a statement on reality through the vagueness of words​, as follows.​

Not​e​ the common sense meaning of one situation being indistinguishable with another​. But​ ``indistinguishable" or ``distinguishable" here has only a well defined mathematical meaning from detection theory, in which the word ``distinguishable" is rarely if ever used for the subjective binary decision which does not refer to the totality of objective reality. In particular, the Leibniz metaphysical principle on the identity of indiscernibles is implicitly used in the ``indistinguishability" justification of security to identify the real situation as the ideal situation with a high probability.​ However, the mathematics does not say that. It would give the probability of each of the two situations only by \textit{assuming} they together exhaust all relevant possibilities, such as the case of deciding between target absent or present. One cannot introduce features to a hypothesis as if it is actual in other considerations after the binary decision. We try to make this clear in the following \textit{example}.

Consider the common problem of radar detection of whether there is an incoming flying object. In a military situation the object of concern could be an enemy airplane, say with or without a warhead. The yes-no target detection problem \textit{cannot} tell by itself whether a warhead is on board. One can't infer the airplane has a warhead because that is hypothesis 1 in the binary detection problem formulation which one puts in \textit{by hand}. Similarly, the occurrence of $\rho_{ideal}$ is an additional unwarranted assumption that one cannot make use of in other problems just from the binary decision problem with $\rho_{ideal}$ as one hypothesis.

The ``distinguishability advantage" justification of operational gurantee from $d$ is thus incorrect in several ways:
\begin{enumerate}[(A)]
\item The a priori probability $p_1$ for the ideal situation cannot be 1/2. Instead it should be 0.
\item The probability of the ideal situation is not $1-d$ from the binary decision. From (4) it is close to 1/2, not 1, for small $d$.
\item The ideal situation cannot be inferred from the binary decision, because it has other features not relevant to the hypotheses testing.
\end{enumerate}

The trace distance $d$ does measure the closeness of the states $\rho_{real}$ and $\rho_{ideal}$, but in a specific mathematical sense only. The \textit{crucial} point is that the security guarantee from $d\leq\epsilon$ has to be deduced from the mathematical statement itself and not by any other meaning involved in the words ``ideal" and ``distinguishability". Exactly the same error of mathematics and reality confusion is made in the failure probability interpretation, to which we now turn.

\section{Failure Probability and Operational Guarantee}

The ``failure probability" interpretation of the trace distance criterion $d$ was first stated and elaborated in [1,2]. A key is called $\epsilon$-secure if $d\leq\epsilon$. It is stated in [1, p.33] that

``an $\epsilon$-secure key can be considered \textit{identical} to an \textit{ideal} (perfect) key- except with probability $\epsilon$" 

(emphasis in original statement).

\noindent And in [2, p.414] it says

``the real and the ideal setting can be considered identical with probability at least $1-\epsilon$".

\noindent This unambiguous but incorrect interpretation is repeated in [9,11] and many other papers; see note [25] of ref [5] for a collection. Some of the many ways in which this interpretation is wrong have been described in [4,5,6,15]. However, the erroneous interpretation has persisted in various ways with no response to the arguments put forth against it. A discussion of the recent view described in [7] in this connection is given in the following.

The original misinterpretation was drawn from Prop. 2.1.1 in [1] which is the same as Lemma 1 in [2]. It is re-stated as Theorem A.6  in [7]. It says that for two random variables $X$ and $Y$ on the same space with probability distributions $P_X$ and $P_Y$ which are marginals of a joint distribution $P_{XY}$, the "coupling inequality"
\begin{equation}
P(X=Y) \geq 1 - \delta (P_X, P_Y)
\end{equation}
is satisfied with equality by a maximizing $P^{\degree}_{XY}$ [18, I.2 and I.5]. Here $X$ is taken to be $K$ and $Y=U$. It was pointed out [4] that the ``there exists" in the mathematical statement cannot validate the conclusion of the misinterpretation $X=Y$ with probability $\delta (P_X, P_Y)$ because there is no reason $P^{\degree}_{XY}$ is in effect. It was further pointed out in [10] that even if it is in effect, the wrong interpretation does not follow. The notation $P( X=Y )$ is a mathematical symbol representing certain probabilities of a joint measurement of $X$ and $Y$, namely,
\begin{equation}
P(X=Y)=\sum_i P(X_i=Y_i)
\end{equation}
where $\{i\}$ are the underlying sample space elements. It does not refer to the probability of some totality of reality concerning $X$ and $Y$.                                   

The basic error here is a confusion of what the mathematics says about reality​. It attributes referents to the symbols which the mathematics itself does not imply. It is exactly of the same nature as the ``distinguishability" confusion discussed ​above​. In App. A.3 of [7], ``failure" is redefined in terms of the joint probability only and the $P^{\degree}_{XY}$ is \textit{assumed} to hold, with the wrong interpretation of $d$ as ``maximum failure probability" maintained as ``an intuitive way of understanding the trace distance". Thus, a mere possibility is elevated to actuality by assuming $P^{\degree}_{XY}$ is in effect, similar to ``distinguishability" in which the situation is taken to be just the real or the ideal one as discussed above.

In fact, the satisfaction of (5) with equality by $P^{\degree}_{XY}$ does \textit{not} imply $X=Y$ with probability $1-\delta (P_X, P_Y)$. Such interpretation is represented mathematically by the existence of a distribution $P'$ such that [6]
\begin{equation}
P_X=(1-\lambda)P_Y+\lambda P'
\end{equation}
for a probability $\lambda$, in this case $\lambda=\delta(P_X,P_Y)$, from the theorem of total probability. Equ (7) is easily shown to hold, from $P'$ being a probability distribution, if and only if
\begin{equation}
\frac{1-\lambda}{N}\leq P(x_i)\leq \lambda+\frac{1-\lambda}{N}
\end{equation}
For large $N$ that $i$ varies up to, equ (8) implies all $P(x_i)$ take essentially the same value around $\lambda$. This condition (8) \textit{cannot} be satisfied by $\lambda=\delta(P_X,U)$ [5]. For any $\lambda$ it implies a uniformity on $P_i$ that does not follow from just a $\delta_E$ level guarantee.

Thus, several errors are committed in the original derivation of the wrong failure probability interpretation, any of which invalidates the derivation:
\begin{enumerate}[(i)]
\item 	There is no reason to expect that the maximizing $P^{\degree}_{KU}$ is in effect.
\item 	The mathematical representation of the failure probability interpretation of $\delta_E$ is not given via (5) and $P^{\degree}_{KU}$. Any joint distribution $P_{KU}$ is irrelevant to such interpretation.
\item 	The correct representation of the failure probability interpretation is given by (7), which cannot hold for $\lambda=\delta_E$, and also not warranted for any $\lambda$ because of (8).
\end{enumerate}

One may conclude that the wrong failure probability interpretation is not valid as well as intuitively misleading. It suggests much stronger security guarantee from $d$ than is actually the case, as one can see directly from the following.

Clearly terminology alone, assuming one wants to stick with the verbally misleading ``indistinguishability", ``failure probability", and ``$\epsilon$-secure", cannot establish security by connotation. One wants mathematically and conceptually correct consequences of the mathematical statement $\delta_E=\epsilon$ to provide security guarantee that has operational and empirical meaning in the real world. In conventional information theoretic security, of which message authentication (necessary in most QKD protocols) is a good example, such operational guarantee are sought in terms of Eve's probability of success in her various attacks on the cryptosystem. In fact there are the following obvious security questions:
\begin{enumerate}[(1)]
\item What is Eve's success probability in estimating the whole key $K$ from her attack during key generation? ​What is it for a subset of $K$?
\item What is the above probability in a known-plaintext attack in which Eve knows part of the data when $K$ is ​used in ``one-time pad" encryption, and hence ​knows ​part of $K$, and gets at the rest of $K$ through ​correlation of bits in $K$?
\item How many bits Eve may get correctly anyway even though the sequence is not correctly estimated in (1)-(2) above?​ This corresponds to a non-uniform a priori distribution of $K$ to Eve and is the bit error rate (BER) issue in ordinary communications.
\item What is the final protocol security when an error correcting code ​or a ​message authentication code is used ​in executing a QKD protocol via ​a ​previously generated QKD key?
\end{enumerate}

What does the failure probability interpretation say about these ​\textit{operational ​s​ecurity ​q​uestions that must be answered} for proper security guarantee? It is clear that whatever $d\leq \epsilon$ may imply, if poor level in the above is not ruled out there is \textit{no} adequate security guarantee​, because Eve may then simply succeed with too high a probability in breaking the system or obtaining significant information despite the declared security from the security criterion​.​ 

These questions are operational because we take probability itself to have operational meaning. However, a theoretical construction from​ probability such as mutual information or statistical distance may not. Its operational meaning has to be explicitly developed mathematically by relating the theoretical quantity to operational probability. In ordinary communications the operational meaning of entropy and mutual information are given through the Shannon coding theorems in terms of the empirical error rate and data rate. For cryptography some operational meaning of mutual information is given in [4], and in [6] for statistical distance.

The failure probability interpretation of $d$ answers (1) that Eve's success probability is bounded by $d$. The correct answer has a uniform $U$ level added [6], which is also given in Lemma A.8 of [7]. Note that the bound can be achieved with equality [4] and hence cannot be improved. The answer to (2) under such interpretation is again $d$, as described in section 5.1 of [7]. However, conditioning on Eve's knowledge from a KPA is not ​then ​taken into account and the answer is incorrect. This is shown by an explicit counter-example in [15] for which Eve's conditional success probability has the maximum value 1 given a specific known portion of $K$. Generally there are average guarantees​ from $d$​ that can be converted to individual probability guarantees via Markov inequality, which greatly weakens the correct ​`​failure probability​"​ guarantee level [6,14]. 

The wrong failure probability interpretation would answer (3) with the computation
\begin{equation}
\mathrm{BER} \leq (1 - d)/2 + d = 1 + d/2                                      
\end{equation}
This unjustified result is more favorable than a correctly derived one for the whole $K$ [10]. For KPA there is no known validly derived BER guarantee from $d$, while (9) would be maintained by the wrong interpretation, which is incorrect from the same counter-example in [15].

While error correction and message authentication are necessary steps in most QKD protocols, their effect on security has never been rigorously quantified [6]. For error correction with open exchange, it appears impossible to quantify Eve's probabilities in points (1)-(3) above with the open information taken into account. If the parity digits of a linear error correcting code (ECC) is covered by a previously generated QKD key, the resulting security is uncertain. According to [7], the security of the final protocol can be obtained by ``universal composition" as
\begin{equation}
d(\rho_{ideal}, \rho_{ecc})\leq d(\rho_{ideal}, \rho_{no \:\:ecc}) + d(\rho_{no\:\:ecc}, \rho_{ecc})
\end{equation}
It seems impossible to meaningfully bound $d(\rho_{no\:\:ecc}, \rho_{ecc}$) with or without the ECC covered by a QKD generated key, and no such result has ever be reported. A brief discussion on message authentication can be found in [6]. Note that operational meaning of any criterion needs to be developed for the purpose under consideration whether it is ``universally composable" or not, as in the case of message authentication.

Thus, for the above basic operational guarantee questions the failure probability interpretation often gives an incorrect answer or no answer at all. Perhaps future work would provide the best correct guarantee of all relevant security problems from the $d$-level. More concretely, even with the wrong interpretation the numerical values of $d$ that can be practically or even theoretically achieved in QKD protocols are troublesome. This numerical situation would be resolved by the new ``failure probability per bit" interpretation, to which we now turn.

\section{Numerical Adequacy of Security Guarantee and Failure \\Probability per Bit}
                                        
​The criterion $d$ applies to a key $K$ generated in a single QKD round. In fact, it would \textit{not} be meaningful to cite a $d$ level without saying how long $K$ is. The numerical adequacy of $d$ depends on such length $|K|$ [6,14]. Indeed, it is possible that under $d\leq\epsilon$ Eve's probability of getting the entire $n$-bit $K$ is given by [4]
\begin{equation}
p_1^E=2^{-n}+d
\end{equation}
Equ(11) achieves the bound in [6,10] and [7, Lemma A.8]. it is clear from (11) that how adequate is the numerical guarantee of a given $d$-level depends on $n=|K|$. For $n=1$, $d\sim10^{-9}$ would be quite good. For $n=10^5$, $d\sim10^{-9}$ is quite poor [6,14,15], as follows.

For $|K|$ in tens of thousands, the best theoretical [9] and experimental value [19] of $d$ is around $10^{-9}$ for single-photon BB84. The best theoretical value​ at vanishing $|K|$ is $10^{-14}$ [9]. Typically 10 QKD rounds are carried out in 1 sec, or ~$\sim 10^6$ rounds per day. The poor security guarantee of the available values are clear even before conversion to individual probability guarantee [6]. After such conversion the possibility of Eve totally breaking about $10^3$ rounds of QKD per day is not ruled out.
See [14] for a brief summary.

A measure $d/|K|$ called ``failure probability per bit" was introduced in [9]. It gives a lower and apparently more desirable security value than $d$ itself, surely, but it is a misleading terminology because it suggests that the bits in $K$ are statistically independent while precisely the bit correlation is the security trouble. A new interpretation of ``failure probability per bit" is given in [7, p.14]:

\begin{enumerate}[(F)]
\item ``For example, if an implementation of a QKD protocol produces a key at a rate of 1 Mbit/s with a failure per bit of $10^{-24}$, then this protocol can be run for the age of the universe and still have an accumulated failure strictly less than 1."
\end{enumerate}

So the failure probability per bit here is $d/l$ where $l$ is the total number of bits generated in all the QKD rounds for the age of the universe at a rate of 1 Mbps, which is $\sim 10^{24}$. In this claim, statistically independent bit leaks seems assumed, which is not correct. More remarkably, the $d$ of a single QKD round output $K$ is taken to apply to $l$ instead of $|K|$. If such argument makes sense there would have been no security problem to begin with, since \textit{any} $d$ value would become arbitrarily small after division by an arbitrarily large number of uses. In particular, the error rate per bit in ordinary communications would be extremely tiny after many uses regardless of how error prone the system is. It is hard not to conclude that such use of failure probability per bit $d/l$ makes no sense.

The actual guarantee in the above quoted numerical example is as follows. Let each QKD round generate $|K|\sim 10^5$ for 10 rounds per second, and take $d\sim 0.1$ to accommodate the quoted values. Then just with the multiple average $d$ (instead of individual) guarantee, already it is not ruled out that for the $10^6$ rounds per day, $10^5$ of them are leaked to Eve with all $10^5$ bits obtained by Eve in each leaked round. This would continue during the whole age of the universe, with $10\%$ of $l$ leaked to Eve constantly. This strongly contradicts the quote (F) that the accumulated ``failure" is strictly less than 1, where ``accumulated failure" is here interpreted as the accumulated bit failure probability as suggested by quote (F), which appears to indicate that the probability of just leaking one bit is strictly less than 1. In particular, if ``accumulated failure" means total compromise of all the $l$ bits, why would it be any security guarantee at all for just a value ``strictly less than 1". Recall that $d$ does not describe the security of all the bits in multiple rounds or gives the failure probability per bit in a given round with output $K$. The rounds are independent but the bits within a single $K$ are not.

\section{Concluding Remark}
                                           Various erroneous interpretations of $d$ were proposed which make QKD appear much more secure than it has been proved. The more significant point is not that mistakes have been made, mistakes are rampant in science and in life. It is that valid criticisms should be addressed explicitly, especially in the area of cryptography in which general security claims cannot be experimentally established.

\end{document}